# Selective Strong-Field Enhancement and Suppression of Ionization with Short Laser Pulses

N. A. Hart,[1,*] J. Strohaber,[2] A. A. Kolomenskii,[1] G. G. Paulus[3], D. Bauer[4] and H. A. Schuessler[1]

[1]*Department of Physics, Texas A&M University, College Station, Texas 77843, USA*

[2]*Department of Physics, Florida A&M University, Tallahassee, Florida 32307, USA*

[3]*Institut fuer Optik und Quantenelektronik, Friedrich-Schiller-Universitaet Jena, Max-Wien-Platz 1, 07743 Jena, Germany*

[4]*Institut fur Physik, Universty of Rostock, 18051 Rostock, Germany*

We experimentally demonstrate robust selective excitation and attenuation of atomic Rydberg level populations in sodium vapor (Na I) using intense laser pulses in the strong field limit ($>10^{12}\,\text{W/cm}^2$). The coherent control of the atomic population and related ionization channels is realized for intensities above the over-the-barrier ionization intensity. A qualitative model predicts that this strong field coherent control arises through the manifestation of a Freeman resonance.

PAC numbers: 32.80.Rm, 33.80.Rv, 42.50.Hz

Practical implementation of resonantly enhanced multiphoton ionization (REMPI) in the strong field limit is complicated by the fact that the intermediate resonances are not static [1]. This is due to dynamic Stark shifts and Stark splittings that satisfy an integer-photon resonance in an intensity dependent manner [2, 3]. For a dipole transition, requiring at least $K$ photons, perturbation theory predicts that the photon absorption rate $W$ is proportional to the intensity $I$ to the $K^{\text{th}}$ power ($W \propto I^K$). However, using Cesium (Cs) atoms, B. Held *et. al.* [4] showed experimentally and C. S. Chang and P. Stehle [5] demonstrated theoretically that a 4-photon ionization can have a $K \equiv Log(W)/Log(I)$ value as low as 1 and as high as 8, depending on the amount of detuning of the laser frequency from the resonance. In this case, $K$ no longer simply represents the multiphoton order.

Freeman *et. al.* [6] discovered that when atomic Rydberg states transiently Stark-shift into resonance, they can be energetically resolved in the photoelectron yield. Observation of the effect requires the pulse duration to be shorter than the time it takes for emitted electrons to exit the laser focus, and the laser bandwidth to be smaller than the spacing between the excited energy levels. This dynamic shift into resonance, commonly referred to as "Freeman resonance", is an important and pervasive phenomenon in strong field physics. This is partly because it satisfies the condition for REMPI and is therefore associated with enhanced coherent ionization [6, 7].

In our experiments, we control the AC Stark shift in atomic sodium by changing the intensity of a femtosecond laser pulse. This allows us to discriminate between multiphoton resonances in a manifold of sodium Rydberg *P* states. By satisfying only one Freeman resonance of the manifold at the peak of the pulse where nonlinear effects are maximum, REMPI through this resonance is selectively enhanced, while other ionization channels are simultaneously suppressed.

Details of our experimental equipment have been described previously [8]. Briefly, our Ti:sapphire laser system produced 57 fs pulses with a center wavelength of 800 nm. A photodiode (PD) detected pulses from a small leakage through the steering mirror M and was used to trigger data-acquisition software (see Figure 1). The power attenuator was a two component device composed of a rotatable half-wave plate that changed the polarization of the laser beam and a fixed polarizing beam-splitter cube that filtered out vertically polarized radiation while passing the horizontal. The sodium oven was located within the vacuum chamber. The oven temperature was maintained at 256° C which produced a Na vapor density of $2\times10^{14}$ cm$^{-3}$ at the exit of the oven and a density of $5\times10^{13}$ cm$^{-3}$ ($10^{-3}$ mbar) at the laser focus ~1.5 cm above the oven. The shortest (transform limited) pulse was created in the focus by adjusting the grating compressor in the laser amplifier. Electrons that were emitted from laser-ionized sodium and traveled down the $\mu$-metal tube of the vacuum chamber in the direction of the electric field of the laser were detected with microchannel plates (MCP) for time-of-flight and energy analysis.

Figure 2 displays three above threshold ionization spectra measured for sodium vapor (Na I) spanning a range of laser intensities from $3.5\times10^{12}$ W/cm$^2$ to $8.8\times10^{12}$ W/cm$^2$. The peaks in the spectra serve as a probe of the atomic level populations during the laser-atom interaction. The location of the peaks on the energy axis were calibrated with sodium atomic energy reference data from the National Institute of Standards and Technology (NIST) [9]. A near resonant two-photon transition between the 3s ground state and 4s excited state favors the excitation of Rydberg P states 5p, 6p and 7p over F states 4f, 5f and 6f [10, 11]. Therefore, one-photon ionization from the 5p and 6p energy levels should provide the most prominent electron peaks at $\approx 0.76$ eV and $\approx 1.03$ eV respectively (see Fig. 2 and Fig. 3).

These peaks, along with all others below 1.55 eV, are shown in the plot one photon ($\hbar\omega \approx 1.55$ eV) in energy away from their respective bound state ($-0.79$ eV for 5p, $-0.51$ eV for 6p and 0 eV for the continuum) and comprise the threshold peak (see Fig. 3). Ionization through the off-resonant 5p state is highly efficient. By fitting a Gaussian to the 5p peaks at 0.76 eV, 2.38 eV and 3.9 eV in the spectra for $I_0 = 5.1\times10^{12}$ W/cm$^2$ and comparing this with the entire electron yield, we estimate that 85% of the photoelectrons are ionized through the 5p intermediate state. It is worth noting that, the Keldysh parameter at the over-the-barrier intensity ($I_{OTB} = 2.79\times10^{12}$ W/cm$^2$) for Na is $\gamma = 3.98$ and indicates that the ionization of sodium with our pulse parameters remains a multiphoton process even after surpassing the predicted saturation intensity [12]. Sustained multiphoton ionization beyond the over-the barrier intensity has been measured previously in lithium [13] and theoretically predicted for potassium [14]. In noble gases, plateau like features in the ATI spectra associated with cutoff energies at 2 times and 10 times the ponderomotive energy of the laser field $U_P$ are important measures for pulse characterization [15, 16]. However, alkalis present a particular challenge for observing such features because of the relatively low intensities needed for ionization. This is due to lower intensities having smaller cutoff energies, which leads to a shortage of peaks below the plateau cutoffs.

Figure 4 displays photoelectron yields of three subpeaks located at $0.76~\text{eV}\,(5p)$, $0.98~\text{eV}\,(6p)$ and $1.17~\text{eV}\,(7p)$. According to lowest order perturbation theory, the threshold subpeaks 5$p$, 6$p$ and 7$p$ should have ionization yields that are proportional to the peak intensity $I_0$ to the $4^{th}$ power ($Y \propto I_0^4$), which on a log-log plot produces a slope of four ($K = 4$). Yet we observe $K_{6p} = 1$ and $K_{5p} = 6$ (Fig. 4).

The difference between the intensity responses of these energy states may be explained by the fact that as intensity increases, the dynamic Stark effect shifts 5$p$ towards a three-photon resonance relative to the ground state, while 6$p$ shifts away from this resonance. 7$p$ also shifts away from the three-photon resonance, but differs from 6$p$ in that 7$p$ lies outside of the three-photon bandwidth of the laser throughout the entire pulse. Therefore, resonant phenomena are not expected to dominate for this higher energy state.

A qualitative understanding of the role that REMPI plays in strong-field ionization may be gained from a simplified model where the multiphoton absorption rate is integrated in a classical two-state rate equation (see Figure 6). Given that the sodium atom is initially in the 3$s$ ground state, this model gives a qualitative estimate of the intensity-dependent population probability for a final excited state $f$ (where $f$ is either 5$p$, 6$p$ or 7$p$) after the pulse has passed. For an instantaneous laser intensity $I$ and a three-photon cross section $\sigma_3$, the three-photon absorption rate is $W = \sigma_3 I^3$. The rate $W$ for the 3$s$ to $f$ transition (3$s$→3$p$→4$s$→$f$) is calculated using lowest order perturbation theory [17]. We introduce the effect of the AC Stark shift $E^{(2)}(t)$ into the model by replacing the static three-photon detuning of the unperturbed atom, $\delta_f^* = 3\omega - \omega_{3s,f}$, with a dynamic detuning, $\delta_f = 3\omega - \omega_{3s,f} - E_f^{(2)}(t)/\hbar$, in the Lorentzian density of states function $\rho(\delta_f, \gamma_f)$ for level $f$ in accordance with Mittleman [18]. The two-photon detuning for the 3$s$→4$s$ transition and the one-photon detuning for the 3$s$→3$p$ transition are $\delta_{4s} = 2\omega - \omega_{3s,4s} - E_{4s}^{(2)}(t)/\hbar$ and $\delta_{3p} = \omega - \omega_{3s,3p}$ respectively.

If we let $\gamma_{3p}$, $\gamma_{4s}$ and $\gamma_f$ equal the decay rates for levels 3p, 4s and f, then the cross section for the four level atom can be expressed as

$$\sigma_3 \approx A\rho(\delta_{3p},\gamma_{3p})\rho(\delta_{4s},\gamma_{4s})\rho(\delta_f,\gamma_f) \quad (1)$$

where A is a constant which does not affect the K value of the ionization process. For arbitrary $\delta$ and $\gamma$ the normalized Lorentzian is

$$\rho(\delta,\gamma) = \frac{1}{\pi}\frac{\gamma/2}{(\delta)^2+(\gamma/2)^2}. \quad (2)$$

To find the population in the final state, we integrate the two-state rate equations

$$\frac{d}{dt}N_{3s} = -W(N_{3s}-N_f), \quad (3)$$

$$\frac{d}{dt}N_f = W(N_{3s}-N_f). \quad (4)$$

Because short pulses have large bandwidths, we average the rate W over the instantaneous bandwidth at each time step of the rate integration through the laser frequency $\omega$. In our case, the line width $\gamma_f$ of $\rho(\delta_f,\gamma_f)$ is much smaller than the laser bandwidth $\Delta\omega$. For this reason $\rho(\delta_f,\gamma_f)$ can be approximated as a delta function which evaluates the rate W at a time-dependent frequency and amplitude of the laser pulse. We refer to this model as resonance sampling (RS). RS is particularly useful in that it isolates the effect of directly populating an excited state associated with REMPI by excluding other excitation mechanisms.

From Fig. 5 it can be seen that inclusion of the AC Stark shift into the rate is able to qualitatively reproduce an increase in $K_{5p}$ and a decrease in $K_{6p}$. The slope $K_{5p}$ of the 5p excitation probability saturates when the resonance condition $\delta_{5p} \approx 0$ is satisfied at approximately $5\times10^{12}$ W/cm$^2$. Contrarily, the slopes of 6p and 7p always decrease from their maximum value of $K=3$, indicating a suppression of the excitation efficiency due to the AC Stark shift.

Two more-general observations can be made from the RS model. Firstly, all higher energy P states (6p, 7p, 8p, etc…) should experience suppressed excitation since they will all shift away from the three-photon resonance by roughly the ponderomotive energy. Moreover, if the intermediate level Stark shifts completely out of the multiphoton bandwidth or is never in resonance during the pulse, then REMPI will not be a dominant ionization mechanism. In this case, intermediate excitation must occur by other means (i.e Rabi oscillations [3]).

Secondly, the greater the static detuning $\delta_f^*$ is above the unperturbed 5p state, the greater the efficiency of 5p excitation. This is because a larger detuning requires a higher intensity to create the Freeman resonance. And since the three-photon excitation rate W has a cubic dependence on intensity, the excitation efficiency is greatly enhanced. This intensity dependence is interesting, because it is the exact opposite of what is expected in the weak field regime.

These trends are also confirmed by a numerical simulation solving the one-dimension (1D) time-dependent Schrödinger equation (TDSE) for a sodium-like atomic well (see Figure 6). The simulation [19] uses a soft-core potential well

$$V(x) = -\frac{ab}{\sqrt{x^2+b^2}} \quad (5)$$

where x is a spatial grid position, $a=0.2411$ and $b=3.6352$. In 1D, even and odd wavefunctions are produced. Therefore to compare results of the simulation with experimental data, we use even states as indicators of S state population and odd states as indicators of P state population. The values for parameters a and b reproduce the eigenvalues for the first three states of sodium 3s, 3p and 4s to two significant figures. However, the higher lying states do not closely reproduce the Rydberg energies. Therefore we chose an arbitrary set of odd intermediate states *f* using a photon energy of $\hbar\omega_s \approx 1.47\text{eV}$, such that one level is slightly below the three-photon resonance, analogous to 6*p* in our experiment. The closest odd states above and below this near $3\hbar\omega_s$ resonance are labeled 7*p* and 5*p* in Fig. 6. The simulation shows a dramatic enhancement of REMPI through a Freeman resonance at 5*p*, as this level shifts into resonance. On the contrary, REMPI through 6*p* is suppressed, showing no intensity dependence over almost an order of magnitude. The 7*p* ionization is non-resonant throughout the entire calculation and does not appear to be involved in REMPI. Here also, selectivity of ionization through an intermediate level is attained at intensities exceeding the over-the-barrier intensity. In agreement with our model, the selective enhancement of REMPI arises from the larger static detuning $\delta_f^*$ relative to higher Rydberg states. In this case, an enhancement of the transition rate *W* correlates with a photon flux which is more than 10 times greater for intensities at the 5*p* dynamic resonance ($\delta_{5p} \approx 0$) than for 6*p* ($\delta_{6p} \approx 0$). Moreover, integrating the electron probability from the 5*p* peaks of each ATI order indicates that this Freeman resonance can saturate the ionization of the atom.

In conclusion, a mechanism is described to account for selective excitation of a single energy state among a manifold of sodium Rydberg energy levels. By Stark shifting the Rydberg states with the laser pulse intensity, we were able to single out a Freeman resonance that selectively enhances REMPI due to its large detuning $\delta_f^*$. REMPI used in this manner was the dominant ionization mechanism since greater than 80% of the experimentally measured electrons were ionized through the target state 5*p*. Simultaneously, REMPI through higher Rydberg states was suppressed by an AC Stark shift away from resonance. Moreover, the coherence of the resonant ionization was maintained and even optimal at intensities exceeding the over-the-barrier intensity of the atom. These results further the understanding of atom-specific REMPI and motivate future applications of strong-field coherent control.

This work was supported by the Robert A. Welch Foundation Grant No. A1546 and the Qatar Foundation under the grant NPRP 5-994-1-172

[1] N.B. Delone, V.P. Kraĭnov, Multiphoton processes in atoms, Springer Science & Business Media, 2000.

[2] K. LaGattuta, Above-threshold ionization of atomic hydrogen via resonant intermediate states, Physical Review A, 47 (1993) 1560.

[3] W. Nicklich, H. Kumpfmüller, H. Walther, X. Tang, H. Xu, P. Lambropoulos, Above-threshold ionization of Cesium under femtosecond laser pulses: New substructure due to strongly coupled bound states, Physical Review Letters, 69 (1992) 3455-3458.

[4] B. Held, G. Mainfray, C. Manus, J. Morellec, F. Sanchez, Resonant Multiphoton Ionization of a Cesium Atomic Beam by a Tunable-Wavelength Q-Switched Neodymium-Glass Laser, Physical Review Letters, 30 (1973) 423-426.

[5] C. Chang, P. Stehle, Theory of Resonant Multiphoton Ionization, Physical Review Letters, 30 (1973) 1283.

[6] R. Freeman, P. Bucksbaum, H. Milchberg, S. Darack, D. Schumacher, M. Geusic, Above-threshold ionization with subpicosecond laser pulses, Physical review letters, 59 (1987) 1092.

[7] G. Gibson, R. Freeman, T. McIlrath, Verification of the dominant role of resonant enhancement in short-pulse multiphoton ionization, Physical review letters, 69 (1992) 1904.

[8] N.A. Hart, J. Strohaber, G. Kaya, N. Kaya, A.A. Kolomenskii, H.A. Schuessler, Intensity-resolved above-threshold ionization of xenon with short laser pulses, Physical Review A, 89 (2014) 053414.

[9] J.E. Sansonetti, Wavelengths, Transition Probabilities, and Energy Levels for the Spectra of Sodium (NaI–NaXI), Journal of Physical and Chemical Reference Data, 37 (2008) 1659-1763.

[10] M. Krug, T. Bayer, M. Wollenhaupt, C. Sarpe-Tudoran, T. Baumert, S. Ivanov, N. Vitanov, Coherent strong-field control of multiple states by a single chirped femtosecond laser pulse, New Journal of Physics, 11 (2009) 105051.

[11] S. Lee, J. Lim, C.Y. Park, J. Ahn, Strong-field quantum control of 2+ 1 photon absorption of atomic sodium, Optics express, 19 (2011) 2266-2277.

[12] R. Potvliege, S. Vučić, Stark-shift induced resonances in multiphoton ionization, Physica Scripta, 74 (2006) C55.

[13] M. Schuricke, G. Zhu, J. Steinmann, K. Simeonidis, I. Ivanov, A. Kheifets, A.N. Grum-Grzhimailo, K. Bartschat, A. Dorn, J. Ullrich, Strong-field ionization of lithium, Physical Review A, 83 (2011) 023413.

[14] F. Morales, M. Richter, S. Patchkovskii, O. Smirnova, Imaging the Kramers–Henneberger atom, Proceedings of the National Academy of Sciences, 108 (2011) 16906-16911.

[15] G.G. Paulus, F. Grasbon, H. Walther, P. Villoresi, M. Nisoli, S. Stagira, E. Priori, S. De Silvestri, Absolute-phase phenomena in photoionization with few-cycle laser pulses, Nature, 414 (2001) 182-184.

[16] D.B. Milošević, G.G. Paulus, D. Bauer, W. Becker, Above-threshold ionization by few-cycle pulses, Journal of Physics B: Atomic, Molecular and Optical Physics, 39 (2006) R203.

[17] R.W. Boyd, Nonlinear optics, Academic press, 2003.

[18] M.H. Mittleman, Multiphoton Ionization, in: Introduction to the Theory of Laser-Atom Interactions, Springer, 1993, pp. 153-231.

[19] G.G. Paulus, W. Nicklich, H. Xu, P. Lambropoulos, H. Walther, Plateau in above threshold ionization spectra, Physical Review Letters, 72 (1994) 2851-2854.

FIGURE CAPTIONS

FIG. 1. (Color online) The experimental apparatus and associated optics. (a) The experimental setup with (M) a steering mirror, halfwave plate (WP), polarizing beam-splitter cube PBC, lens (L), sodium oven and power meter (PM) are shown along the beam path. A photodiode triggers the acquisition of electron counts from the chevron microchannel plates (MCP).

FIG. 2. (Color online) Experimentally measured above threshold ionization (ATI) spectra at laser intensities $3.5 \times 10^{12}$ W/cm$^2$ (blue, lowest), $4.9 \times 10^{12}$ W/cm$^2$ (red, middle) and $8.8 \times 10^{12}$ W/cm$^2$ (black, highest). The data shows the first three photon orders of ATI peaks whose maximum values are located at 0.76eV, 2.4eV and 3.9eV. Substructures in the threshold peak owing to ionization through Rydberg states are labeled 5$p$, 6$p$ and 7$p$. While the substructure labeled 3$d$ belongs to the second order peak.

FIG. 3. (Color online) Energy level diagram for atomic sodium. The continuum shown is a surface plot of 30 experimentally measured ATI spectra presented in this paper. The ATI peaks are horizontal streaks and the laser intensity for each measurement is increasing towards the right with a range of $3.5 \times 10^{12}$ W/cm$^2$ to $1.0 \times 10^{13}$ W/cm$^2$. Each diagonal arrow represents a photon absorption of approximately $\hbar\omega \approx 1.55$ eV in energy. Two possible quantum paths are shown whereby photon absorption produces ionization peaks.

FIG. 4. (Color online) Experimentally measured electron yields for ionization from the 5$p$ (blue circles), 6$p$ (purple triangles) and 7$p$ (green diamonds) energy levels as a function of laser intensity. The maximum value of the 5$p$, 6$p$ and 7$p$ peaks from the same data set as Figure 2 and Figure 3 are plotted. In the plot, the red solid line has a slope of $K = 6$, the grey dashed line is for $K = 1$ and the green dotted line has $K = 2$.

FIG. 5. (Color online) Results of the resonance sampling (RS) model for three-photon absorption by 5$p$, 6$p$ and 7$p$ levels as a function of the pulse peak intensity. The curves for 6$p$ and 7$p$ have been shifted vertically. 5$p$ excitation

has a slope $K_{5p} = 8$ (red solid line), while 6p and 7p both have a slope $K \approx -0.5$ (gray dashed and green dotted lines respectively). It should be noted that at low intensities ($<10^{11}$ W/cm$^2$), all three levels have a slope $K = 3$. The relatively low excitation probabilities for 6p and 7p suggest that REMPI is not the dominant mechanism for ionization through these states over the modeled intensity range.

FIG. 6. (Color online) Numerically simulated electron yields for ionization through 5p (blue circles), 6p (purple triangles) and 7p (green diamonds) energy levels as a function of laser intensity. As in Fig. 4, the maximum value of each peak is plotted. The red solid line refers to $K_{5p} = 14$, the grey dashed line is the result for $K_{6p} = 0$ and the green dotted line refers to $K_{7p} = 4$.

FIGURES

Fig. 1

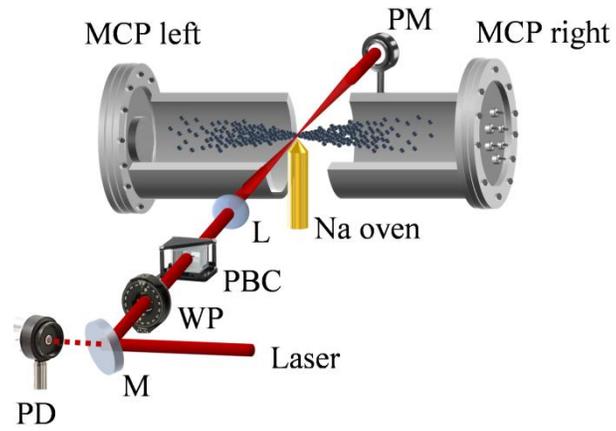

Fig. 2

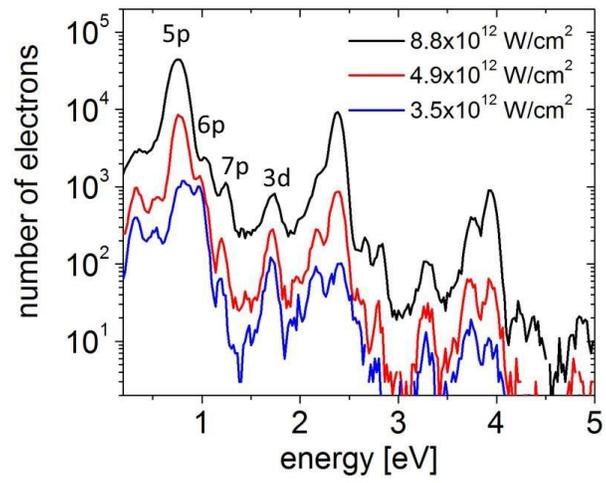

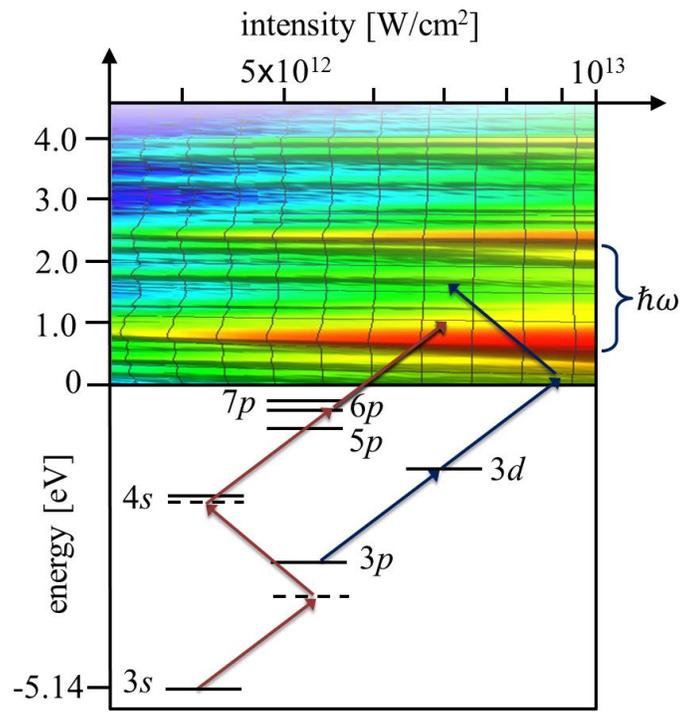

Fig. 3

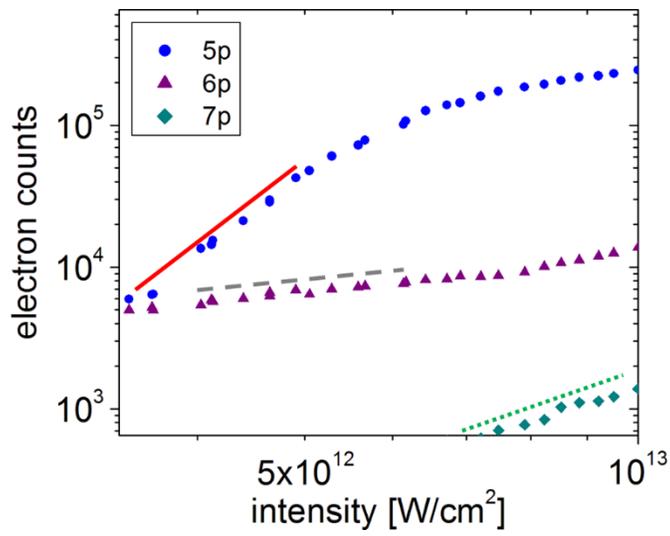

Fig. 4

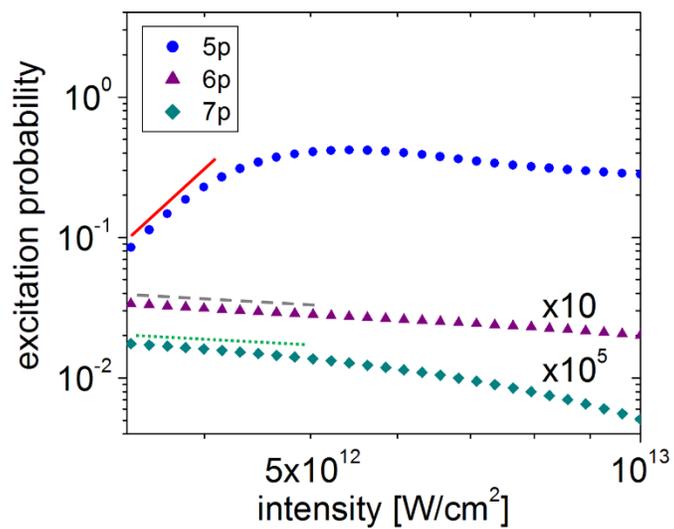

Fig. 5

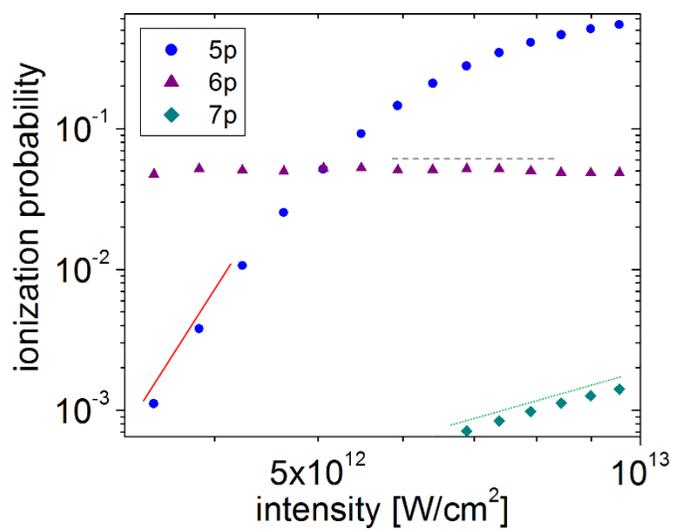

Fig. 6